\UseRawInputEncoding
\documentclass[a4paper]{article}
\usepackage{amsfonts}
\usepackage{graphicx}
\usepackage{amssymb}
\usepackage{amsmath}
\usepackage{mathrsfs}
\usepackage{latexsym}
\usepackage{geometry}
\usepackage{authblk}
\usepackage{verbatim}
\usepackage{subfigure}
\usepackage{hyperref}
\usepackage{epstopdf}
\usepackage{color}
\usepackage{cite}

\newcommand{\bean}{\begin{eqnarray}}
\newcommand{\eean}{\end{eqnarray}}
\newcommand{\eqs}[1]{Eqs. (\ref{#1})}
\newcommand{\eq}[1]{Eq. (\ref{#1})}
\newcommand{\meq}[1]{(\ref{#1})}
\newcommand{\fig}[1]{Fig. \ref{#1}}

\newcommand{\tab}[1]{Tab. \ref{table1}}

\newcommand{\oh}{\frac{1}{2}}

\newcommand{\bea}{\begin{eqnarray*}}
\newcommand{\eea}{\end{eqnarray*}}
\newcommand{\grad}{\nabla}

\newcommand{\eqn}{&=&}
\newcommand{\non}{\nonumber \\}

\begin{document}
\title{Testing the weak cosmic censorship conjecture for extremal magnetized Kerr-Newman black holes}
\author{Yunjiao Gao\thanks{201821140015@mail.bnu.edu.cn}}
\author{Sijie Gao\thanks{Corresponding author: sijie@bnu.edu.cn}}
\affil[1]{Department of Physics, Beijing Normal University, Beijing , China,100875}

\maketitle
\begin{abstract}

We test the  weak cosmic censorship conjecture for magnetized Kerr-Newman spacetime via the method of injecting a test particle. Hence, we need to know how the black hole's parameters change when a test particle enters the horizon. This was an unresolved issue for non-asymptotically flat spacetimes since there are ambiguities on the energies of black holes and particles. We find a novel approach to solve the problem. We start with the  `` physical process version" of the first law, which relates the particle's parameters with the change in the area of the black hole. By comparing this first law with the usual first law of black hole thermodynamics, we redefine the particle's energy such that the energy can match the mass parameter of the black hole.  Then, we show that the horizon of the extremal magnetized Kerr-Newman black hole could be destroyed after a charged test particle falls in, which leads to  a possible violation of the weak cosmic censorship conjecture. We also find that the allowed parameter range for this process is very small, which indicates that  after the self-force and radiation effects are taken into account, the weak cosmic censorship conjecture could still be valid. In contrast to the case where the magnetic field is absent, the particle cannot be released at infinity to destroy the horizon. And in the case of a weak magnetic field, the releasing point becomes closer to the horizon as the magnetic field increases. This indicates that the magnetic field makes the violation of the cosmic censorship more difficult. Finally, by applying our new method to  Kerr-Newman-dS (AdS) black holes, which are well-known non-asymptotically flat spacetimes, we obtain the expression of the particle's energy which matches the black hole's mass parameter.
\end{abstract}

\section{Introduction}\label{sec1}
 Penrose proposed the ``weak cosmic censorship conjecture (WCCC)" in 1969 \cite{penrose1969gravitational}. It states that gravitational collapse always results in a black hole instead of a naked singularity. In other words, singularities formed in a gravitational collapse are always hidden in a black hole.
 Whether the WCCC is correct is still one of the most important open problems in general relativity. In 1974, Wald proposed a gedanken experiment to verify the WCCC \cite{wald1974gedanken}. It was shown that the horizon of an extremal  Kerr-Newman (KN)  black hole can not be destroyed by injecting a test particle into the  black hole.  In 1999, Hubeny extended the gedanken experiment to nearly extremal Reissner-Nordstrom (RN) black holes \cite{hubeny1999overcharging}and proved that the horizon of the black hole can be destroyed by charged particles if the particle's parameters are carefully chosen.  In recent years, people have constructed a large number of gedanken experiments to test the WCCC and found more apparent violations \cite{gao2013destroying,zhang2014testing,jacobson2009overspinning,
 Revelar:2017sem,deFelice:2001wj,Cardoso:2015xtj,Yang:2020iat,Feng:2020tyc}. These experiments ignore the influence of the test particle on the geometry of spacetime. A break through was made by Sorce and Wald in 2017\cite{sorce2017gedanken}. They obtained the second-order perturbation inequality about the mass, angular momentum, and charge of the KN black hole. In the new version of the gedanken experiment, the second-order perturbation just saves the WCCC.  This approach  has been applied to some other stationary black holes and the WCCC are always valid
\cite{Ge:2017vun,jiang2020testing,ning2019gedanken,jiang2020new,
wang2020examining,jiang2020investigating,Zhang:2020txy}. Recently, the WCCC and the second law of black hole thermodynamics have been discussed in the extended phase space \cite{Zeng:2019aao,Zeng:2019huf,Zeng:2019hux}. Moreover, some studies suggest that black holes and naked singularities could be distinguished observationally by their gravitational lensing characteristics, allowing the WCCC to be tested by future observations \cite{Virbhadra:2002ju,Virbhadra:2007kw}.

In this paper, we are interested in testing the WCCC for a magnetized black hole. It has been discovered that there is a strong magnetic field near supermassive black holes in the active galactic nuclei\cite{Zhang:2005yq}, indicating that we need to take into account  the  magnetic field when studying the WCCC. When the magnetic field is weak, people can ignore its influence on the geometry of spacetime. For example, when a near-extremal Kerr black hole surrounded by a weak magnetic field, the spacetime can still be described by the Kerr metric\cite{shaymatov2015destroying}. In this case, it has been shown that it is possible to turn the  black hole into a naked singularity by injecting a test particle\cite{shaymatov2015destroying}. However, for strong magnetic fields, its backreaction to the spacetime geometry cannot be ignored. Therefore, it is necessary to find an exact solution to the Einstein's equation describing a magnetized black hole. In fact,  by using the Ernst technique\cite{1976JMP....17..182E,ernst1976black,hiscock1981black,ernst1968new},  the magnetized Kerr-Newman (MKN) solution has  been found in the literature\cite{aliev1989exact,dokuchaev1987black}, which satisfies the sourceless Einstein-Maxwell's equation. In these works, the metric components are  given only in the linear order in $B$, where $B$ is the strength of the magnetic field. Following the same method, we first obtain the  exact MKN solution. Our solution could reduce to that in  \cite{aliev1989exact,dokuchaev1987black} by taking the linear order in $B$. We then use this solution to test the WCCC.

 Due to the magnetic field, the MKN spacetime is not asymptotically flat. This causes difficulties when testing the WCCC because both the mass of the black hole and the energy of the test particle cannot be uniquely defined. How to define the test particle's energy such that it can match the energy change of a black hole is an open issue. To solve this problem, we revisit the ``physical process version" of the first law\cite{wald1994quantum,gao2001physical}. Unlike the usual first law of black hole thermodynamics, this first law only involves local properties of the black hole and the global nature of the spacetime is irrelevant. Then by comparing the two first laws, we give the definition of the particle's energy which can match the mass parameter of the MKN black hole. By applying this new definition, we  test the WCCC for extremal MKN black holes and find the parameter ranges which lead to possible violation of the WCCC. Finally, we apply the same argument to  Kerr-Newman-dS (AdS) black holes, which are typical non-asymptotically flat spacetimes, and obtain the expression of the particle's energy matching the mass of black hole.

 The rest of this paper is organized as follows. In section \ref{sec2}, we review the Ernst equation and derive the exact MKN metric as well as the corresponding electromagnetic field. In section \ref{sec3}, we test whether the extremal MKN black hole can be destroyed by a charged particle. In section \ref{sec4}, we derive the energy of a test particle in the Kerr-Newman-dS (AdS) black hole. Conclusions and discussions are given in section \ref{sec5}. In appendix \ref{appc}, we prove that the second law of black hole thermodynamics holds if the null energy condition is satisfied.  We adopt the geometric units $c=G=1$ through this paper.

\section{Review of magnetized Kerr-Newman black holes}\label{sec2}
 Consider a general stationary axisymmetric solution satisfying Einstein-Maxwell equations described by the metric
\begin{equation}
\mathrm{d}s^2=f^{-1}\left(\rho^2\mathrm{d}t^2-e^{-2\gamma}\left(\frac{\mathrm{d}r^2}{\Delta_r}+\frac{\mathrm{d}x^2}{\Delta_x}\right)\right)-f\left(\mathrm{d}\phi-\omega\mathrm{d}t\right)^2,\label{sgxkl}
\end{equation}
where $\rho=\sqrt{\Delta_{r}\Delta_x}$, $\Delta_r=r^2+a^2+Q^2-2Mr$, $\Delta_x=1-x^2(x= \cos\theta)$, and $f$, $\omega$, $e^{-2\gamma}$ are functions of $r$ and $x$. The constants $M$, $Q$ and $a$ will be interpreted as the mass, charge and angular momentum per mass later.
The corresponding vector potential takes the form
\begin{equation}
A_{a}=A_t{dt}_{a}+A_\phi{d\phi}_{a}\label{gsckvf}.
\end{equation}
Ernst \cite{ernst1968new} shows that the solution \meq{sgxkl} and \meq{gsckvf} can de described by a pair of complex potential, namely the electromagnetic potential $\Phi$ and the gravitational potential $\varepsilon$:
\begin{align}
\Phi &=A_\phi+iB_\phi\label{ineq5},\\
\varepsilon &=\left(f-\Phi\Phi^{\star}\right)+i\varphi\label{ineq6},
\end{align}
with \cite{ernst1968new,aliev1989exact}
\begin{align}
 -iD B_\phi &=\rho^{-1}f \left(D A_t+\omega D A_\phi\right)\label{ineq7},\\
D\varphi &=-i\rho^{-1}f^2D\omega+i\left(\Phi^{\star}D\Phi-\Phi D\Phi^{\star}\right)\label{ineq8},
\end{align}
where \cite{siahaan2017destroying}
\begin{equation}
D=
\frac{2\sqrt{\Delta_r \Delta_x}\left(\left(r-M\right)\partial_r-x\partial_x\right)+i\left(2x\Delta_r\partial_r+\Delta_x\left(2r-2M\right)\partial_x\right)}{2x^2\Delta_r+\left(r-M\right)\Delta_x\left(2r-2M\right)}\label{ineq21}.
\end{equation}
Making use of above definitions, Einstein-Maxwell equations can be written  in the symmetric form
\begin{align}
f\nabla^2 \varepsilon &=\left(\nabla\varepsilon+2\Phi^{\star}\nabla\Phi\right)\cdot\nabla\varepsilon\label{ineq2},\\
f\nabla^2 \Phi &=\left(\nabla\varepsilon+2\Phi^{\star}\nabla\Phi\right)\cdot\nabla\Phi\label{ineq3},
\end{align}
where $\nabla$ is the three-dimensional divergence operator.
\eqs{ineq2} and \meq{ineq3} are called Ernst equations.

It can be shown
\cite{kinnersley1973generation}  that \eqs{ineq2} and \meq{ineq3} are invariant under  $SU(2,1)$  transformations. By applying this transformation, we can generate a new set of solutions $(\tilde{\Phi}, \tilde{\varepsilon })$ from the old solutions $(\Phi,  \varepsilon)$. The new solutions also satisfy Ernst equations \meq{ineq2} and \meq{ineq3}. Thus, it follows from \eqs{sgxkl} and \meq{gsckvf} that
\begin{equation}
\mathrm{d}\tilde s^2=\tilde{f}^{-1}\left(\rho^2\mathrm{d}t^2-e^{-2\gamma}\left(\frac{\mathrm{d}r^2}{\Delta_r}+\frac{\mathrm{d}x^2}{\Delta_x}\right)\right)-\tilde{f}\left(\mathrm{d}\phi-\tilde{\omega}\mathrm{d}t\right)^2,\label{cklbf}
\end{equation}
\begin{equation}
\tilde{A}_{a}=\tilde{A}_t{dt}_{a}+\tilde{A}_\phi{d\phi}_{a}.
\end{equation}
 $\tilde{\Phi}$, $\tilde{\varepsilon}$, $\tilde{B}_{\phi}$ and $\tilde{\varphi}$ of the new solution still satisfy the \eqs{ineq5}-\meq{ineq8} with $f$ replaced by $\tilde{f}$.

Next, we apply the $SU(2,1)$ transformation to the KN metric to generate a new solution, which describes a black hole immersed in a uniform magnetic field, i.e., the MKN black hole. We start with the KN black hole, where the metric functions in \eq{sgxkl} are given by
\bean
f\eqn -\frac{\left(r^2+a^2\right)^2-\Delta_r a^2\Delta_x}{r^2+a^2x^2}\Delta_x\label{ineq15},\\
\omega\eqn \frac{a\left(2Mr-Q^2\right)}{\left(r^2+a^2\right)^2-\Delta_r a^2\Delta_x}\label{ineq17}, \\
e^{-2\gamma}\eqn \left(\left(r^2+a^2\right)^2-\Delta_r a^2\Delta_x\right)\Delta_x.
\eean
The corresponding vector potential is
\begin{equation}
A_{a}=-\frac{Q r}{r^2+a^2 x^2}{\mathrm{d}t}_{a}+\frac{Q r a \Delta_x}{r^2+a^2 x^2}{\mathrm{d}\phi}_{a}\label{ineq12}.
\end{equation}
Making using of \eqs{ineq5} and \meq{ineq6}, we can derive
\begin{equation}
\Phi=\frac{a-i r x}{r+i a x}Q,\label{zcmk}
\end{equation}
\begin{equation}
\varepsilon=-\left(r^2+a^2-a\frac{2Ma+i\left(2Mr-Q^2\right)x}{r+i a x}\right)\Delta_x-\left(4Ma+i Q^2x\right)\frac{a-i r x}{r+i a x},\label{bkms}
\end{equation}
which is same as the result of \cite{ernst1976black}.

The corresponding transformation in the $SU(2,1)$ group that can generate a solution to MKN black hole  is \cite{astorino2013embedding,ernst1976black,hiscock1981black,siahaan2016magnetized}
\begin{align}
\tilde{\varepsilon} &=\Lambda^{-1}\varepsilon,\label{ineabbb}\\
\tilde{\Phi} &=\Lambda^{-1}\left(\Phi-\frac{1}{2}B\varepsilon\right)\label{ineqqq},
\end{align}
where
\begin{equation}
\Lambda=1+B\Phi-\frac{1}{4}B^2\varepsilon\label{ineqq},
\end{equation}
and the parameter $B$ is a constant, which represents the strength of the external  magnetic field.
 Under the transformation,
\begin{equation}
\tilde{f}=\left({\rm Re}\,\varepsilon+\Phi\Phi^{\star}\right)\vert\Lambda\vert^{-2}=f\vert\Lambda\vert^{-2},\label{gsklcd}
\end{equation}
where $\vert\Lambda\vert^{2}=\Lambda\Lambda^{\star}$ and ${\rm Re}\,\varepsilon=f-\Phi\Phi^{\star}$ have been used.

Applying the transformation given in \eqs{ineabbb}, \meq{ineqqq} and \meq{ineqq} to the electromagnetic potential \meq{zcmk} and the gravitational potential \meq{bkms} of the KN metric, we can derive  $\tilde{\omega}$, $\tilde{A}_\phi$, $\tilde{A_t}$(Please see App.\,\ref{appa} for the concrete expressions).

Replacing $\tilde{f}$ in the line element \meq{cklbf} with \eq{gsklcd}, the MKN metric becomes
\begin{equation}
\mathrm{d}s^2=-\frac{f}{\vert\Lambda\vert^2}\left(\mathrm{d}\phi-\tilde{\omega}\mathrm{d}t\right)^2+\frac{\vert\Lambda\vert^2}{f}\left(\rho^2\mathrm{d}t^2-e^{-2\gamma}\left(\frac{\mathrm{d}r^2}{\Delta_r}+\frac{\mathrm{d}x^2}{\Delta_x}\right)\right).\label{zcfsx}
\end{equation}

However, the solution has a conical singularity since $\phi$ ranges from $0$ to $2\pi\vert\Lambda\left(x=1\right)\vert^2$ \cite{hiscock1981black}, and some singular stress-energy tensors will be generated on the right side of Einstein's equation. The singularity can be removed by a rescaling  $\phi\rightarrow\tilde{\phi}=\frac{\phi}{\vert\Lambda\left(x=1\right)\vert^2}$ \cite{aliev1989magnetized,aliev1989exact,astorino2015microscopic}. Thus, the final  solution is
\begin{equation}
\mathrm{d}s^2=-\frac{f}{\vert\Lambda\vert^2}\left(\vert\Lambda\left(x=1\right)\vert^2\mathrm{d}{\tilde{\phi}}-\tilde{\omega}\mathrm{d}t\right)^2
+\frac{\vert\Lambda\vert^2}{f}\left(\rho^2\mathrm{d}t^2-e^{-2\gamma}
\left(\frac{\mathrm{d}r^2}{\Delta_r}+\frac{\mathrm{d}x^2}{\Delta_x}\right)\right)\label{inabcd},
\end{equation}
where
\begin{equation}
\vert\Lambda\left(x=1\right)\vert^2=\left(-a B^2 M - B Q\right)^2 + \left(1 + \left(B^2 Q^2\right)/4\right)^2\label{1ab}.
\end{equation}

\begin{equation}
\tilde{A}_{\tilde{\phi}}=\vert\Lambda\left(x=1\right)\vert^2\left(\tilde{A}_{\phi0}+\tilde{A}_\phi\right),\label{vsboc}
\end{equation}
and $\tilde{A}_{\phi0}$ is added to make $\tilde{A}_{\tilde{\phi}}=0$ at $x=1$\cite{astorino2015microscopic,siahaan2017destroying}.

We have verified that the MKN metric satisfies the sourceless Einstein-Maxwell equations.
When $a=0$, $\tilde{\omega}$ reduces to the result of the magnetized Reissner-Nordstrom  black hole\cite{astorino2015microscopic}. When $Q=0$, it reduces to the magnetized Kerr black hole \cite{astorino2015magnetised}. When $B=0$, it reduces to the KN black hole. Note that our $\tilde{A}_t$ determined by \eq{ineq34} differs from that in \cite{astorino2015magnetised} by a constant $B M \left[ 2a M + 3 a^2 B Q+ 2 B Q \left(-M^2 + Q^2\right)\right]/a^2$, which doesnot affect subsequent calculations. In the linear approximation, our $\tilde{A}_t$ can return to that in  Ref.~\cite{dokuchaev1987black}. The metric of  Ref.\cite{gibbons2013ergoregions} is consistent with our metric after coordinate and gauge transformations and removing the conical singularity, which still does not affect subsequent calculations.

The magnetic field takes the form
\begin{equation}
B_{a}=-\frac{1}{\vert\Lambda\vert}\, ^\star F_{ab} \left(\frac{\partial}{\partial t}\right)^{b},
\end{equation}
where $\vert\Lambda\vert$ is introduced to ensure that the observer is normalized. Since the expression of the magnetic field  is lengthy, we only consider the background case, i.e., $M=a=Q=0$, where the magnetic field is given by
\begin{equation}
B_{a}=\frac{1}{\vert\Lambda\vert}(-x B  (\mathrm{d}r)_{a} -r B  (\mathrm{d}x)_{a})= -\frac{1}{\vert\Lambda\vert}B (\mathrm{d}z)_{a},
\end{equation}
or
\begin{equation}
B^{a}=-B g^{zz}\frac{1}{\vert\Lambda\vert} \left(\frac{\partial}{\partial z}\right)^{a}=-B \vert\Lambda\vert^{-3} \left(\frac{\partial}{\partial z}\right)^{a},
\end{equation}
with $\vert\Lambda\vert=1+\frac{B^2 \rho^2}{4}$. It is manifest that the magnetic field in the  background spacetime is uniform along the $z$ direction.

\section{Using test particles to destroy the extremal magnetized Kerr-Newman black hole
}\label{sec3}

In order to obtain the maximum energy needed for a particle to destroy the black hole, we must identify  the mass, electric charge  and angular momentum of the MKN spacetime. Unfortunately, since the magnetized spacetime is not asymptotically flat, the energy  cannot be uniquely defined.
In the following, we shall start with the definitions in  Ref.\cite{astorino2016mass}. The mass, angular momentum and electric charge are given by
\bean
\tilde{M}\eqn \left[M^2 + 2  B J Q  +
 B^2 \left(2  J^2 +\frac{3 M^2 Q^2}{2} - Q^4\right)+ B^3 J Q \left(2 M^2 -\frac{3 Q^2}{2}\right)\right.\non
 && +\left.
  B^4  \left(\frac{Q^4 M^2}{16} +   M^2J^2 - \frac{Q^2J^2}{2}\right)\right]^{\frac{1}{2}}\label{ineq42}, \\
\tilde{J}\eqn J  - B Q^3 - \frac{3}{2}  B^2 J Q^2- \frac{1}{4} B^3 Q \left(8 J^2  + Q^4\right) -
 \frac{1}{16}  B^4 J \left(16 J^2 + 3 Q^4\right)\label{ineq43}, \\
\tilde{Q}\eqn Q+2 J B   - \frac{B^2 Q^3}{4}\label{ineq44},
\eean
where, $J=M a$.

One can check that $\tilde{M}$, $\tilde{J}$, and $\tilde{Q}$ satisfy the following relationship\cite{astorino2016mass}
\begin{equation}
  \tilde{M}^2 = \frac{A_{H}}{16 \pi} + \frac{\tilde{Q}^2}{2} + \frac{\pi (\tilde{Q}^4 + 4 \tilde{J}^2)}{A_{H}}\label{xcne}.
\end{equation}
And the desired first law is
\begin{equation}
\delta\tilde{M}=\alpha\left( \frac{\kappa}{8 \pi}\delta A_{H}+(\Phi_{H}-\Phi_{int})\delta\tilde{Q}+(\Omega_{H}-\Omega_{int})\delta\tilde{J}\right)
\label{innnn2},
\end{equation}
where,
 $\alpha=M/\tilde{M}$
\bean
\Omega_{int}\eqn \frac{ - 4 B Q (4 + B^2 (-4 M^2 + Q^2))+3 a B^4 M (4 M^2 - Q^2)}{M (16 + 16 a^2 B^4 M^2 + 32 a B^3 M Q + 24 B^2 Q^2 + B^4 Q^4)}, \\
\Phi_{int}\eqn \frac{-32 a M B + (-48 M^2 Q - 32 Q^3) B^2 + (-48 a M^3 - 132 a M Q^2) B^3}{2 M (16 + 16 a^2 B^4 M^2 + 32 a B^3 M Q + 24 B^2 Q^2 + B^4 Q^4)}\nonumber\\&+&\frac{(-128 a^2 M^2 Q - 24 M^2 Q^3 + 8 Q^5) B^4 + (-32 a^3 M^3 -
    36 a M^3 Q^2 + 15 a M Q^4) B^5}{2 M (16 + 16 a^2 B^4 M^2 + 32 a B^3 M Q + 24 B^2 Q^2 + B^4 Q^4)}\nonumber\\&+&\frac{(-8 a^2 M^4 Q + 6 a^2 M^2 Q^3 + M^2 Q^5) B^6}{2 M (16 + 16 a^2 B^4 M^2 + 32 a B^3 M Q + 24 B^2 Q^2 + B^4 Q^4)}.
\eean

Now we consider throwing a charged particle into the magnetized Kerr-Newman black hole and see whether the horizon can be destroyed.
The Lagrangian of the charged particle can be written as
\begin{equation}
\mathscr{L}=\frac{1}{2}mg_{\mu\nu}\frac{\mathrm{d}x^\mu}{\mathrm{d}\tau}\frac{\mathrm{d}x^\nu}{\mathrm{d}\tau}+q A_\mu\frac{\mathrm{d}x^\mu}{\mathrm{d}\tau}\label{apuef},
\end{equation}
where $m$ and $q$ are defined as the mass and charge of the test particle, respectively.
The two Killing vector fields, $\frac{\partial}{\partial{t}}$ and $\frac{\partial}{\partial{\phi}}$,  in the MKN spacetime yield the conserved energy and angular momentum along the worldline:
\bean
E\eqn -\frac{\partial{\mathscr{L}}}{\partial\dot{t}}=-\left(m g_{tt}\dot{t}+m g_{t\phi}\dot{\phi}+q A_t\right)\label{innm1}, \\
L\eqn \frac{\partial{\mathscr{L}}}{\partial\dot{\phi}}=m g_{\phi\phi}\dot{\phi}+m g_{t\phi}\dot{t}+q A_{\phi}\label{innnn1}.
\eean
In addition, the four-velocity of the particle, $u^\mu=\frac{dx^\mu}{d\tau}$, gives the constraint
\bean
g_{\mu\nu}u^\mu u^\nu=-1  \,. \label{umunu}
\eean
Through   \eqs{innm1}-\meq{umunu}, we obtain
\begin{equation}
\begin{aligned}
E=&-\left(q A_t+\frac{g_{t\phi}}{g_{\phi\phi}}\left(L-q A_{\phi}\right)\right)\\&+\frac{1}{g_{\phi\phi}}\sqrt{\left(g_{t\phi}^2-g_{t t}g_{\phi\phi}\right)\left(\left(L-q A_{\phi}\right)^2+g_{\phi\phi}m^2\left(g_{r r}\dot{r}^2+g_{xx}\dot{x}^2+1\right)\right)} \,. \label{eex}
\end{aligned}
\end{equation}
 Here, we have taken the plus sign in front of the square root, as $u^\mu$ is a future-directed timelike vector, which means
$\dot{t}=\frac{\mathrm{d}t}{\mathrm{d}\tau}>0$.

It can be seen from the MKN metric \meq{inabcd} that the extremal condition is  $M^2=a^2+Q^2$. If the particle entering the black hole can make the final parameters satisfy $M^2<a^2+Q^2$, the horizon of the black hole is then destroyed. In asymptotically flat spacetimes, we can simply add the energy of the particle to the mass of the black hole, which is unique, i.e., the ADM mass.
 However, it is obviously not the case for the MKN metric. So we need to find out how a particle changes the mass parameter of the black hole. For this purpose, we use the physical process version of the first law and define the energy of the particle as (please refer to Appendix  \ref{appb} for detailed derivation)
\begin{equation}
E'=\alpha E-\alpha\Omega_{int}L-\alpha \Phi_{int} q\label{kk1},
\end{equation}
which satisfies
\begin{equation}
E'=\alpha\left( \frac{\kappa}{8 \pi}\delta A_{H}+(\Phi_{H}-\Phi_{int})q+(\Omega_{H}-\Omega_{int})L\right).\label{kpev}
\end{equation}
Comparing \eq{innnn2} with \eq{kpev}, we find that $E'$, $q$, and $L$ can be regard as $\delta\tilde{M}$, $\delta\tilde{Q}$, and $\delta\tilde{J}$ respectively. Therefore, after the particle crosses the horizon, the mass of the black hole changes from $\tilde M$ to  $\tilde M +E'$.

By \eqs{eex} and \meq{kk1}, we have
\begin{equation}
\begin{aligned}
E'=&-\alpha\left(q \left(A_t+\Phi_{int}\right)+L\Omega_{int}+\frac{g_{t\phi}}{g_{\phi\phi}}\left(L-q A_{\phi}\right)\right)\\&+\frac{\alpha}{g_{\phi\phi}}\sqrt{\left(g_{t\phi}^2-g_{t t}g_{\phi\phi}\right)\left(\left(L-q A_{\phi}\right)^2+g_{\phi\phi}m^2\left(g_{r r}\dot{r}^2+g_{xx}\dot{x}^2+1\right)\right)}\label{ineq40}\,,
\end{aligned}
\end{equation}
which leads to the inequality
\begin{equation}
E'>-\alpha\left(q (A_t+\Phi_{int})+L\Omega_{int}+\frac{g_{t\phi}}{g_{\phi\phi}}(L-q A_{\phi})\right).
\end{equation}
 At the event horizon $r=r_+$, we have  $g_{t\phi}^2-g_{t t}g_{\phi\phi}=0$. Thus, if the particle can cross the horizon, it must satisfy
\begin{equation}
E'>E'_{\rm min}\equiv-\alpha\left(q (A_t+\Phi_{int})+L\Omega_{int}+\frac{g_{t\phi}}{g_{\phi\phi}}(L-q A_{\phi})\right)\Big|_{r=r_{+}}\label{ineq41}.
\end{equation}

On the other hand, it follows from
 Eqs.~\eqref{ineq42}- \eqref{ineq44} that
\begin{equation}
\tilde{M}^4-\tilde{Q}^2\tilde{M}^2-\tilde{J}^2=M^2 \left(M^2-a^2-Q^2\right)\vert\Lambda(x=1)\vert^4\label{efghij}.
\end{equation}
So the extremal condition can be rewritten as
\bean
\tilde{M}^4-\tilde{Q}^2\tilde{M}^2-\tilde{J}^2=0\,.
\eean
 If the final state satisfies
\begin{equation}
\left(\tilde{M}+E'\right)^4-\left(\tilde{Q}+q\right)^2\left(\tilde{M}
+E'\right)^2-\left(\tilde{J}+L\right)^2<0\label{fh}.
\end{equation}
the horizon will disappear.
To solve inequality \meq{fh}, we shall follow the analysis in Ref.\cite{gao2013destroying}. First, we define
\begin{equation}
W=\left(\tilde{M}+E'\right)^2.
\end{equation}
Then \eq{fh} becomes
\begin{equation}
W^2-\left(\tilde{Q}+q\right)^2 W-\left(\tilde{J}+L\right)^2<0,
\end{equation}
which means $W_{1}<W<W_{2}$ with
\begin{equation}
W_{1,2}=\frac{\left(\tilde{Q}+q\right)^2 \mp \sqrt{\left(\tilde{Q}+q\right)^4 + 4 \left(\tilde{J}+L\right)^2}}{2}.\label{w1w2}
\end{equation}
Apparently, $W_{1}<0$, and $W_{2}>0$.
From $W=\left(\tilde{M}+E'\right)^2$ and $W_{1}<0<W<W_{2}$, we have $\left(\tilde{M}+E'\right)^2<W_{2}$. Therefore
\begin{equation}
E'<\sqrt{W_{2}}-\tilde{M}. \label{wm}
\end{equation}
Subsitituting $W_{2}$ in \eq{w1w2} into \eq{wm}, we obtain
\begin{equation}
\begin{aligned}
E'<\sqrt{W_{2}}-\tilde{M}=\frac{\sqrt{2 \left(\tilde{Q} + q\right)^2 + 2 \sqrt{\left(\tilde{Q} + q\right)^4 + 4 \left(\tilde{J} + L\right)^2}}}{2}
- \tilde{M}\equiv E'_{\rm max}\label{ineq46},
 \end{aligned}
\end{equation}
\eqs{ineq41} and \meq{ineq46} give the constraints of a particle which can reach the black hole and destroy its horizon. If the solution exists, it must satisfy
\begin{equation}
\Delta E' \equiv E'_{\rm max}-E'_{\rm min}>0.\label{ineq48}
\end{equation}

To find out whether such solutions exist, we define
\begin{equation}
W_{3}=\left(\tilde{M}+E'_{\rm min}\right)^2\label{in10}.
\end{equation}
\eq{ineq46} leads to
\begin{equation}
 W_{2}=\left(\tilde{M}+E'_{\rm max}\right)^2.\label{in11}
 \end{equation}
Therefore, \eq{ineq48} can be equivalently written as
\begin{equation}
\begin{aligned}
s&\equiv W_{2}-W_{3}\\&=\frac{\left(\tilde{Q}+q\right)^2 + \sqrt{\left(\tilde{Q}+q\right)^4 + 4 \left(\tilde{J}+L\right)^2}}{2}-\left(\tilde{M}+E'_{\rm min}\right)^2
>0,\label{nxm}
\end{aligned}
\end{equation}
The exact expression of $s$ \meq{nxm} is very complicated. Obviously, the particle's parameters $q$ and $L$ are small quantities compared to the black hole's parameters.  We shall focus on the case where the magnetic field is weak. We first expand $s$ to the second order of $q$ and $L$ and then expand it to the second-order of $B$. Thus, \eq{nxm}  becomes
 \begin{equation}
\frac{1}{\left(a^2 + M^2\right)^3}\left(C_1 q^2 +C_2 q L +C_3 L^2 \right)>0\label{in1},
\end{equation}
with
\begin{subequations}
\begin{align}
 C_1=&2 a^2 M^2 \left(3 M^2 - a^2\right) + 4  M a Q \left(2 a^4+ 2 a^2 M^2 - 3  Q^4\right)B\nonumber \\&+2  \left(8 a^8 + 4 a^6 Q^2- 42 a^4 Q^4 - 36 a^2 Q^6 +3 Q^8\right)B^2,\\
 C_2=&-2  M a Q \left(3 M^2 - a^2\right) - 2 \left(4 a^6 + 18 a^4 Q^2 + 12 a^2 Q^4 - 3 Q^6\right)B\nonumber\\&-  3  M a Q \left(16 a^4 + 7 a^2 Q^2 - \frac{43}{2} Q^4\right)B^2 ,\\
 C_3=&M^2 \left(M^2 - 3 a^2\right)+2  M  a Q \left(5 M^2 - 3 a^2\right)B  + \frac{8 a^6 + 66 a^4 Q^2 + 51 a^2 Q^4 - 13 Q^6}{2} B^2.
\end{align}
\end{subequations}
When $B=0$, \eq{in1} goes back to the result of  Ref.~\cite{gao2013destroying}.
Note that the  coefficients are approximately in the order of magnitude
\begin{subequations}
\begin{align}
C_{1}&\sim M^6+B M^7+B^2 M^8,\\
C_{2}&\sim M^5+B M^6+B^2 M^7,\\
C_{3}&\sim M^4+B M^5+B^2 M^6.
\end{align}
\end{subequations}
Therefore, only when  $B\ll 1/M$, one could drop the higher orders of $B$. Hereinafter, we shall only consider $B$ in this range.
Since the leading order of the coefficient of $q^2$ in \eq{in1} is positive, when $q$ is non-zero and $L$ is small enough, \eq{in1} always has solutions. Therefore, we have demonstrated that the violation of the cosmic censorship is possible.

\begin{figure}
  \centering
  \includegraphics[width = 12 cm]{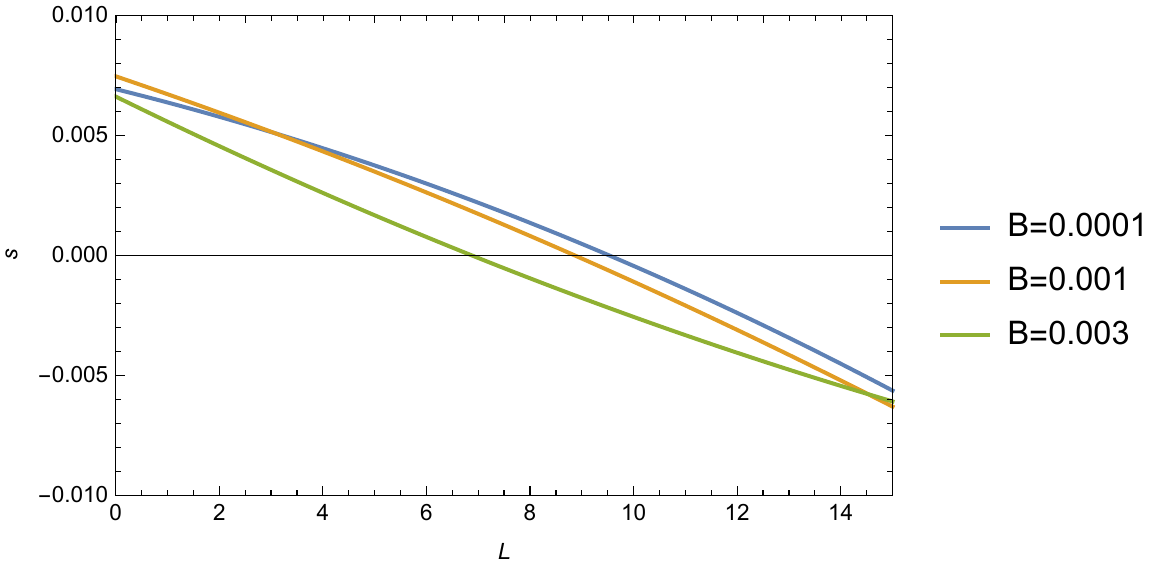}
  \caption{Plot of $s$ varying with $L$. The corresponding parameters are $M = 100$, $a= 80$, $Q=60$, $q=0.1$, and $B = \{ 0.0001, 0.001, 0.003\}$. When $L$ is small, $s$ is positive, meaning that the horizon could be destroyed.}\label{fig3.1}
\end{figure}

In order to verify the above conclusion, we take  $M=100$, $a=80$, $Q=60$, $q=0.1$, $B = \{ 0.0001, 0.001, 0.003\}$. Choosing $q=0.1$ is to satisfy $q<<Q$.~\fig{fig3.1} shows that when $B<<\frac{1}{M}=0.01$, as long as $L$ is small enough,  $s$ is always positive.

However, to realize the violation, a fine tuning on the particle's energy is required.
According to \eqs{in10} and \meq{in11}, we have $s\equiv W_{2}-W_{3}\sim 2\tilde{M}\Delta E'$. \eq{in1}
implies that when $B<<\frac{1}{M}$, $s\equiv W_{2}-W_{3}$ is of order $q^2$ or $ \frac{L^2}{M^2}$. For weak magnetic fields, \eq{ineq42} shows  $\tilde{M}\sim M$. Thus,  $\Delta E'$ is of order $\frac{q^2}{M}$ or $\frac{L^2}{M ^3}$. For illustration, we take $M=100$, $a=80$, $Q=60$, $q=0.1$, $L=5$, $B=0.0001$. Then we can obtain $\tilde{M}=100.487\sim M=100 $ and $E'_{\rm min}=0.0619454<E'<0.061964=E'_{\rm max}$. Therefore, $\Delta E'=1.86342\times10^{-5}$, which is consistent with  $\frac{q^2}{M}=1\times10^{-4}$ and $\frac{L^2}{M^3}=2.5\times10^{- 5}$.

To see the particle's trajectory more clearly, we consider the particle moving on the equatorial plane $\theta=\pi/2$ ($x=0$). The radial behavior is determined by the effective potential which is defined by
\bean
V_{eff}=-\oh \dot r^2 \,.
\eean
By solving \eq{ineq40} for $\dot r^2$, we obtain
\begin{equation}
\begin{aligned}
V_{eff}&=-\frac{-g_{t\phi}^2 m^2 (1 + g_{xx} \dot{x}^2) \alpha^2 +
 gtt (L^2 - 2 A_{\phi} L q + A_{\phi}^2 q^2 +
    g_{\phi\phi} m^2 (1 + g_{xx}
    \dot{x}^2)) \alpha^2}{2 g_{rr} (g_{t\phi}^2 - g_{tt} g_{\phi\phi}) m^2 \alpha^2}\\&-\frac{2 g_{t\phi} (L -
    A_{\phi} q) \alpha (A_{t} q \alpha + E' +
    q \alpha \Phi_{int} + L \alpha \Omega_{int}) +
 g_{\phi\phi}(A_{t} q \alpha + E' +
    q \alpha \Phi_{int} + L \alpha \Omega_{int})^2}{2 g_{rr} (g_{t\phi}^2 - gtt g_{\phi\phi}) m^2 \alpha^2}\,,\label{iii}
\end{aligned}
\end{equation}
where we have used $\dot x=0$ on the equatorial plane.

\begin{figure}[!hbt]
  \centering
  \includegraphics[width = 7 cm, height = 5 cm]{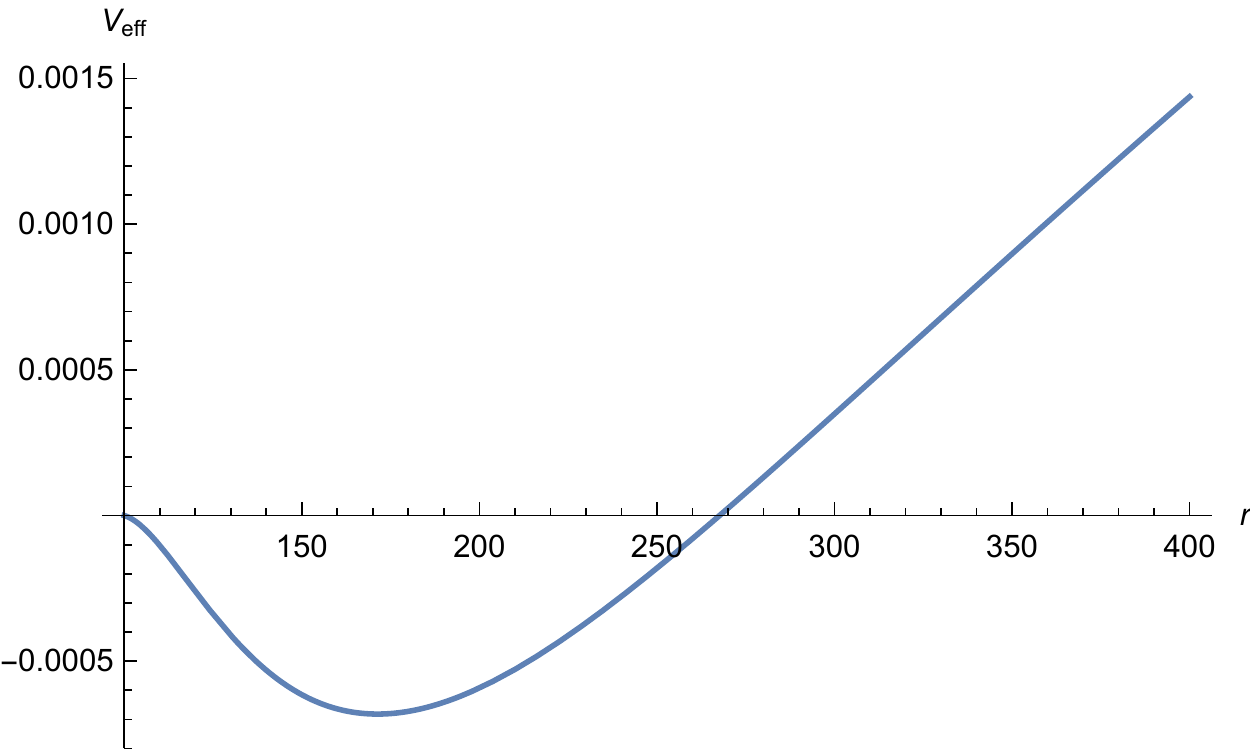}
  \caption{ Plot of $V_{eff}(r)$. The parameters are $M=100$, $a=80$, $Q=60$, $q=0.1$, $L=5$, and $B=0.0001$. The horizon is located at $r=100$. $V_{eff}<0$ at the horizon. So the particle released somewhere outside the horizon, such as $r=200$, can fall into the black hole.} \label{fig3.5}
\end{figure}

We still take $M=100$, $a=80$, $Q=60$, $q=0.1$, $L=5$, $B=0.0001$ as above, and $m=0.0619499$, $E' = 0.061959$, which is within the allowed range( $E'_{\rm min}=0.0619454<E'<0.061964=E'_{\rm max}$). Then we can deduce $E=0.0611496$, and plot $V_{eff}(r)$, as shown in \fig{fig3.5}. For the extremal black hole, $r_{+}=M=100$. When $r=r_{+}=100$, $V_{eff}=-6.44492\times10^{-8}<0$, which means that the particle can reach the horizon from somewhere outside the horizon and could destroy the black hole.

To see the role of the magnetic field, we first consider $B=0$. It is not difficult to check that  $V_{eff}\to(1- E'^{2}/ m^2)/2$ at infinity, which means that, when  $E'> m$, the particle can be released at infinity and go all the way to the horizon.

\begin{figure}[!hbt]
  \centering
  \includegraphics[width = 10 cm, height = 5 cm]{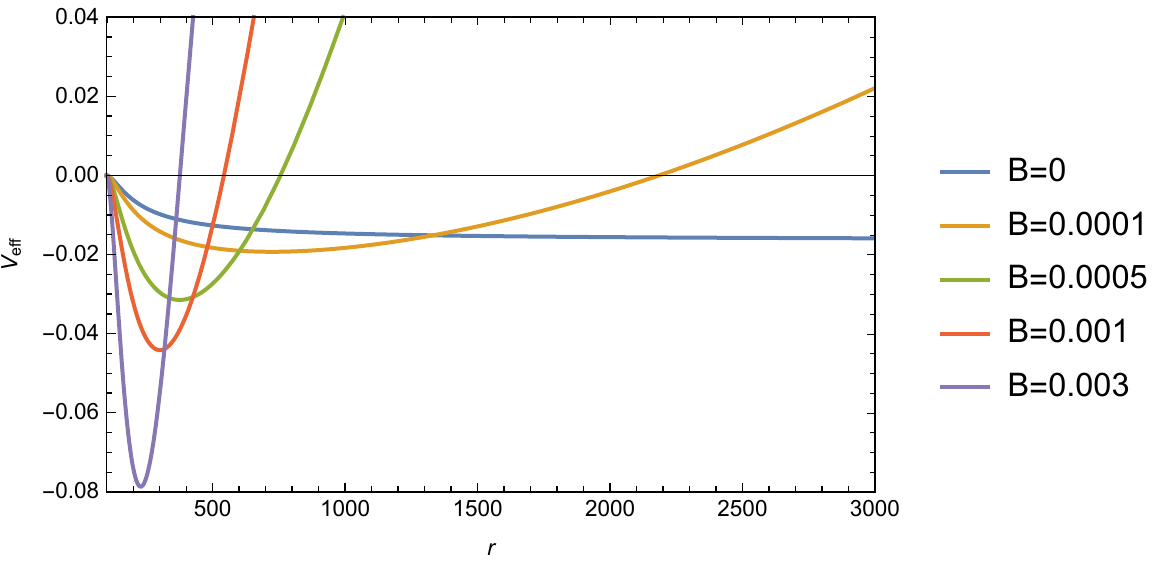}
  \caption{ Plot of $V_{eff}(r)$. The parameters are chosen as $M=100$, $a=80$, $Q=60$, $q=0.1$, $L=5$, and $m=0.06$. The horizon is located at $r=100$. The regions of $V_{eff}<0$ are permitted to release the test particle such that the particle could reach the horizon. It shows that, in the case of a weak magnetic field, the stronger the magnetic field is, the closer we have to release the particle from the event horizon.} \label{fig3.6}
\end{figure}

 When $B \neq 0$, we expand the $V_{eff}$ at infinity as
\bean
  V_{eff}|_{r\to \infty}=\frac{32 (-2 B L + 4 q + 2 a B^3 M q Q + 3 B^2 q Q^2)^2}{B^2 m^2 (16 + 16 a^2 B^4 M^2 + 32 a B^3 M Q + 24 B^2 Q^2 +
   B^4 Q^4)^2 }\frac{1}{r^2}+O(\frac{1}{r^3}).\label{fg}
\eean
Since the leading order of \eq{fg} is positive,  we always have $V_{eff}\to 0^{+}$ at infinity. Therefore, in the presence of the magnetic field $B$, the particle cannot fall into a black hole from infinity to destroy the event horizon. Moreover, as demonstrated in \fig{fig3.6}, the releasing point of the particle ($V_{eff}$ just turns to be negative) becomes closer to the horizon with the increase of $B$. These results suggest that the magnetic field makes the violation of the WCCC more difficult.

\section{Energy of test particles in Kerr-Newman-dS(AdS) black holes }\label{sec4}
When the cosmological constant is taken into account, spacetimes are not asymptotically flat. It is not so clear how the black hole changes when a particle falls in. In this section, we shall apply the method in the previous section to define the energy of particle in KN-dS(AdS) black holes.

The metric of the KN-dS(AdS) black hole is \cite{Gwak:2018tmy,caldarelli2000thermodynamics}
\begin{equation}
d s^{2}=-\frac{\Delta_{r}}{\rho^{2}}\left[d t-\frac{a \sin ^{2} \theta}{\Xi} d \phi\right]^{2}+\frac{\rho^{2}}{\Delta_{r}} d r^{2}+\frac{\rho^{2}}{\Delta_{\theta}} d \theta^{2}+\frac{\Delta_{\theta} \sin ^{2} \theta}{\rho^{2}}\left[a d t-\frac{r^{2}+a^{2}}{\Xi} d \phi\right]^{2},
\end{equation}
where
\begin{equation}
\rho^{2}=r^{2}+a^{2} \cos ^{2} \theta, \quad \Xi=1+\frac{\Lambda a^{2}}{3},
\end{equation}
\begin{equation}
\Delta_{r}=\left(r^{2}+a^{2}\right)\left(1-\frac{\Lambda r^{2}}{3}\right)-2 M r+Q^{2}, \quad \Delta_{\theta}=1+\frac{\Lambda a^{2}}{3} \cos ^{2} \theta.
\end{equation}
The mass $\tilde{M}$, angular momentum $\tilde{J}$ and charge $\tilde{Q}$ are given by
\begin{equation}
  \tilde{M}=\frac{M}{\Xi^2},\quad \tilde{J}=\tilde{M}a=\frac{M a}{\Xi^2},\quad \tilde{Q}=\frac{Q}{\Xi}\,.
\end{equation}
Define the angular velocity and electric potential on the horizon $r=r_+$ as
\bean
\Omega_{H}\eqn \frac{a \Xi}{a^2+r_{+}^2}\,, \\
\Phi_{H}\eqn \frac{q r_{+}}{a^2+r_{+}^2}\,.
\eean
One can check that the first law of thermodynamics takes the form\cite{Gwak:2018tmy,caldarelli2000thermodynamics}
\begin{equation}
\delta\tilde{M}= \frac{\kappa}{8 \pi} \delta A_{H}+(\Omega_{H}-\Omega_{\infty})\delta\tilde{J}+\Phi_{H}
\delta\tilde{Q}\label{first},
\end{equation}
where $\Omega_{\infty}=\frac{a \Lambda}{3}$  represents the angular velocity  at infinity.

Now, we consider a test particle falling into the black hole. The particle can be treated as a perturbation described by the  stress-energy tensor $\Delta T_{ab}$. Then
 the Einstein's equation becomes
\begin{equation}
G_{ab}+\Lambda g_{ab}=R_{ab}-\frac{1}{2}g_{ab}R+\Lambda g_{ab}=8\pi \left(T_{ab}^{EM}+\Delta T_{ab}\right)\label{yuasd}.
\end{equation}
By following almost the same derivation in appendix \ref{appb}, we can obtain the same physical process version of the first law \footnote{  Although the first law derived in the appendix \ref{appb} is based on the standard Einstein's equations without the cosmological constant,  one can see that the $\Lambda$ term in \eq{yuasd} does not appear in the final result.   }
\begin{equation}
 E=\frac{\kappa\delta A_H}{8 \pi} +\Omega_{H}L+\Phi_{H}q\label{afirst}.
\end{equation}
Then we define
\begin{equation}
E'=E-\Omega_{\infty}L=\frac{\kappa\delta A_H}{8 \pi} +(\Omega_{H}-\Omega_{\infty})L+\Phi_{H}q\label{hhhjjk}.
\end{equation}
By comparing \eq{hhhjjk} with \eq{first}, we see that $E'$ is just the energy that could be added to $\tilde M$.

It is worth noting that Gwak defined the particle's energy in KN-dS(AdS) spacetimes \cite{Gwak:2018tmy}, denoted by $E_0$, in order to prevent the violation of the second law of black hole thermodynamics.
It is straightforward to show that $E_0$ is exactly the same as $E'$ in our \eq{hhhjjk}.

\section{Conclusions}\label{sec5}
In this paper, by using the Ernst equation and the SU(2,1) invariance, we first drive the complete MKN metric and the corresponding electromagnetic field. We have verified that this metric satisfies sourceless Einstein-Maxwell's equations.  Since the MKN spacetime is not asymptotically flat, there are ambiguities in the mass of  black hole and the energy of  particle. To solve this problem, we compare the ``physical process version'' of the first law with the ordinary first law of black hole thermodynamics and redefine the particle's energy such that it matches the mass of the black hole. By applying this energy,  we have found that when   $L$ is small enough and $B<<\frac{1}{M}$, the particle could enter and destroy the horizon of the extremal MKN black hole, making a possible violation of the WCCC. We also showed that the stronger the magnetic field is, the harder it becomes to destroy the horizon. On the other hand, we  found that the allowed energy range is very small, in the order of $\frac{q^2}{M}$ or $\frac{L^2}{M^3 }$. Therefore, if the self-energy and self-radiation effects are taken into account, the WCCC could be saved. Finally, we applied the same strategy to the KN-dS(AdS) black hole and found the expression of the particle's energy that matches the mass of the black hole. Our approach can be further applied to other non-asymptotically flat spacetimes and tell how the mass parameter of black hole changes when a particle falls in, which plays a crucial role in testing the WCCC.

\section*{Acknowledgements}
 The authors thank Marco Astorino for useful discussions. This research was supported by NSFC Grants No. 11775022 and 11873044  .

\appendix
\section{ The specific form of the magnetized Kerr-Newman metric}\label{appa}
Since the specific expression of $\tilde{\omega}$ in the MKN metric \meq{inabcd} is lengthy, we list it in this appendix, namely
\begin{equation}
\tilde{\omega}=\frac{\omega_{0}+\omega_{1}B+\omega_{2}B^2+\omega_{3}B^3+\omega_{4}B^4}{H},\label{zmvs}
\end{equation}
where
\begin{align}
H&=(r^2 + a^2)^2 - \Delta_{r} a^2 \Delta_x \nonumber \\&=r^4 + a^4 x^2 + a^2 \left(Q^2 \left(-1 + x^2\right) + r \left(2 M + r - 2 M x^2 + r x^2\right)\right),\\
\omega_{0}&=-a\left(Q^2 - 2 M r\right),\\
\omega_{1}&=-2 Q r \left(a^2 + r^2\right),\\
\omega_{2}&=-\frac{3}{2}a  Q^2 \left(Q^2 x^2 - 2 M r x^2 + a^2
\left(1 + x^2\right) + r^2 \left(1 + x^2\right)\right),\\
\omega_{3}&=\frac{1}{2}Qr^3 \left(Q^2 x^2 - 2 M r x^2 + r^2 \left(1 + x^2\right)\right)-\frac{1}{2} Q a^4 \left(r - 3 r x^2 + M \left(2 + 4 x^2\right)\right)\nonumber\\& +
   \frac{1}{2} Q a^2\left(Q^2\left(-2 r - 4 M x^2 + 3 r x^2\right) +
       2 r \left(M r + 4 M^2 x^2 - 5 M r x^2 + 2 r^2 x^2\right)\right),\\
\omega_{4}&= \frac{1}{4}a  M^2 r^4 x^2 \left(-6 + x^2\right) -
      \frac{1}{16}a Q^2 \left(Q^2 + r^2\right)x^2 \left(Q^2 x^2 - 3 r^2 \left(-2 + x^2\right)\right)\nonumber\\& -\frac{1}{8}a
       M \left(-Q^4 r x^4 + 4 Q^2 r^3 x^2 \left(-3 + x^2\right) +
         r^5 \left(-3 - 6 x^2 + x^4\right)\right) \nonumber\\&-\frac{1}{16}
      a^5 \left(Q^2 \left(1 + x^2 + x^4\right) + 4 M^2 \left(1 + 3 x^2 + x^4\right)-
         2 M r \left(-1 + 6 x^2 + 3 x^4\right)\right)\nonumber\\&+\frac{1}{16}
      a^3
         4 M r \left(2 M^2 x^2 \left(3 + x^2\right) + M r \left(3 - 9 x^2 - 4 x^4\right) +
            r^2 \left(1 + 6 x^2 + x^4\right)\right)\nonumber\\& -\frac{1}{16}
      a^3
         Q^2 \left(4 M^2 x^2 \left(3 + x^2\right) + r^2 \left(1 + 7 x^2 - 2 x^4\right) -
            2 M r \left(-3 + 7 x^2 + 4 x^4\right)\right)\nonumber\\& -\frac{1}{16}
      a^3 Q^4 \left(1 + x^2 + 2 x^4\right).
\end{align}

We rewrite \eq{ineq7} as
\begin{equation}
D X=A_\phi D \omega-i\rho f^{-1}D B_\phi\label{ineqxdf},
\end{equation}
where $X=A_t+\omega A_\phi$. Hence,
\begin{equation}
\tilde{A}_t=\tilde{X}-\tilde{\omega}\tilde{A}_\phi\label{ineq34}.
\end{equation}

According to \eq{ineq34}, we can get the $t$ component ($\tilde{A}_t$) corresponding to the electromagnetic four-vector potential of the MKN metric, where the specific expression of $\tilde{X}$ is as follows:
\begin{equation}
\tilde{X}=\frac{X_{0}+X_{1}B+X_{2}B^2+X_{3}B^3}{H},\label{xvsnf}
\end{equation}
with
\begin{align}
X_{0}&=-Q r \left(a^2 + r^2\right),\\
X_{1}&=-\frac{3}{2}a  Q^2 \left(Q^2 x^2 - 2 M r x^2 + a^2 \left(1 + x^2\right) + r^2 \left(1 + x^2\right)\right),\\
X_{2}&=\frac{3}{4}Q r^3 \left(Q^2 x^2 - 2 M r x^2 + r^2 \left(1 + x^2\right)\right) -\frac{3}{4}Q
   a^4 \left(r - 3 r x^2 + M \left(2 + 4 x^2\right)\right) \nonumber\\&+\frac{3}{4}Q
   a^2 \left(Q^2 \left(-2 r - 4 M x^2 + 3 r x^2\right)+
      2 r \left(M r + 4 M^2 x^2 - 5 M r x^2 + 2 r^2 x^2\right)\right),\\
X_{3}&=\frac{1}{2}a   M^2 r^4 x^2 \left(-6 + x^2\right) -\frac{1}{8}a
    Q^2 \left(Q^2 + r^2\right) x^2 \left(Q^2 x^2 - 3 r^2 \left(-2 + x^2\right)\right)\nonumber\\& -\frac{1}{4}a
     M \left(-Q^4 r x^4 + 4 Q^2 r^3 x^2 \left(-3 + x^2\right)  +
       r^5 \left(-3 - 6 x^2 + x^4\right)\right)\nonumber\\& -\frac{1}{8}
    a^5 \left(Q^2 \left(1 + x^2 + x^4\right) + 4 M^2 \left(1 + 3 x^2 + x^4\right) -
       2 M r \left(-1 + 6 x^2 + 3 x^4\right)\right)\nonumber\\& +\frac{1}{8}
    a^3
       4 M r \left(2 M^2 x^2 \left(3 + x^2\right) + M r \left(3 - 9 x^2 - 4 x^4\right) +
          r^2 \left(1 + 6 x^2 + x^4\right)\right) \nonumber\\& -\frac{1}{8}
    a^3
       Q^2 \left(4 M^2 x^2 \left(3 + x^2\right) + r^2 \left(1 + 7 x^2 - 2 x^4\right) -
          2 M r \left(-3 + 7 x^2 + 4 x^4\right)\right)\nonumber\\& -\frac{1}{8}
    a^3 Q^4 \left(1 + x^2 + 2 x^4\right).
\end{align}
The expression of $ \tilde{A}_\phi$ is as follows \cite{gibbons2013ergoregions}:
\begin{equation}
  \tilde{A}_\phi = \frac{(\chi_0 + \chi_1 B + \chi_2 B^2 + \chi_3 B^3)}{\left(r^2 + a^2 x^2\right) \vert\Lambda\vert^2},\label{bsfkl}
\end{equation}
where
\begin{align}
  \chi_0 &= a Q r \left(1-x^2\right)\nonumber\\
  \chi_1 &=1/2 \left(H \left(1-x^2\right) + 3 Q^2 \left(a^2 + r^2 x^2\right)\right)\nonumber\\
  \chi_2 & =
 \frac{3}{4} a Q r \left(r^2 + a^2\right) \left(1-x^2\right)^2 +
  \frac{3}{2} a Q M \left(r^2 \left(3 - x^2\right) x^2 + a^2 \left(1 + x^2\right)\right) - 3/4 a Q^3 r \left(1-x^2\right) x^2\nonumber\\
  \chi_3 &=\frac{1}{8} \left(r^2 + a^2 x^2\right) \left(r^2 + a^2\right)^2 \left(1-x^2\right)^2 +  \frac{1}{2} a^2  M r \left(r^2 + a^2\right) \left(1-x^2\right)^3 -
 \frac{1}{2} a^2 Q^2 M r \left(5 - x^2\right)\nonumber\\& \left(1-x^2\right) x^2 +
  \frac{1}{2} a^2  M^2 \left(r^2 \left(3 - x^2\right)^2 x^2 + a^2 \left(1 + x^2\right)^2\right) +
 \frac{1}{4} Q^2 \left(r^2 + a^2\right)\nonumber\\& \left(r^2 + a^2 + a^2 \left(1-x^2\right)\right) \left(1-x^2\right) x^2 +
 \frac{1}{8} Q^4 \left(r^2 x^2 + a^2 \left(2 - x^2\right)^2\right) x^2\nonumber\\
 \vert\Lambda\vert^2&= 1 +\frac{\Lambda_1 B + \Lambda_2 B^2 + \Lambda_3 B^3 + \Lambda_4 B^4}{r^2 + a^2 x^2}\nonumber\\
  \Lambda_1&=2 a Q r \left(1-x^2\right)\nonumber\\
  \Lambda_2&=\frac{1}{2} \left(\left(r^2 + a^2\right)^2 - \Delta_{r} a^2 \left(1-x^2\right)\right) \left(1-x^2\right) +
 \frac{3}{2} Q^2 \left(a^2 + r^2 x^2\right)\nonumber\\
 \Lambda_3 &=\frac{-Q a \Delta_{r}}{2 r} \left(r^2 \left(3 - x^2\right) x^2 + a^2 \left(1 + x^2\right)\right) +\frac{
 a Q \left(r^2 + a^2\right)^2 \left(1 + x^2\right)}{2 r} \nonumber\\& +\frac{
 Q^3 a \left(\left(2 r^2 + a^2\right) x^2 + a^2\right)}{2 r} \nonumber\\
 \Lambda_4&=\frac{1}{16} \left(r^2 + a^2\right)^2 \left(r^2 + a^2 x^2\right)  \left(1-x^2\right)^2 + \frac{1}{4} M a^2 r \left(r^2 + a^2\right) \left(1-x^2\right)^3 \nonumber\\&+
 \frac{1}{4} M a^2 Q^2  r \left(x^2 - 5\right) \left(1-x^2\right) x^2 +
 \frac{1}{4} M^2  a^2 \left(r^2 \left(x^2 - 3\right)^2 x^2 + a^2 \left(1 + x^2\right)^2\right) \nonumber\\&+
 \frac{1}{8} Q^2 \left(r^2 + a^2\right) \left(r^2 + a^2 + a^2 \left(1-x^2\right)\right) \left(1-x^2\right) x^2 +
 \frac{1}{16} Q^4 \left(r^2 x^2 + a^2 \left(2-x^2\right)^2\right) x^2\nonumber\\
\end{align}
Putting the above results together, we can obtain $\tilde A_t$ in \eq{ineq34}.

\section{ The energy of the test particle in the magnetized Kerr-Newman black hole  }\label{appb}
In this section, we will define the particle's energy outside the MKN black hole such that it can match the black hole's mass parameter \cite{wald1994quantum}. Consider an extended test particle with $u^a$ being its four-velocity.
We define the energy-momentum tensor $\Delta T_{ab}$  and 4-current density $j^{a}$ by
\begin{equation}
\Delta T_{ab}=\mu u_{a}u_{b},\quad\quad j_{a}=\rho u_{a},
\end{equation}
where $\mu$ is the  mass density and $\rho$ is the  charged density.
We denote the energy, angular momentum, and charge of the test particle by $E$, $L$, and $q$, respectively. Suppose $t^a$ is a timelike Killing vector field. Then the energy of a test particle with mass $m$ and charge $q$ is given by\cite{gao2013destroying}
\begin{equation}
\begin{aligned}
E&=-(m u_{a}+q A_{a})t^{a}\\&=-\int_{\Sigma}\left(\mu u_{a}+\rho A_{a}\right) t^{a} \,^{3}\!\epsilon \\&
=\int_{\Sigma}\left(u^{b}\Delta T_{ab}+u^{b}j_{b} A_{a}\right) t^{a}  \,^{3}\!\epsilon\\&=\int_{\Sigma}\left(\Delta T_{ab}+A_{a}j_{b}\right) t^{a}u^{b}  \,^{3}\!\epsilon,\label{ases}
\end{aligned}
\end{equation}
where  the integral is performed on a three-dimensional spacelike hypersurface $\Sigma$ with $u^{a}$  being its normal  and  $^{3}\!\epsilon$ being its volume element.
Similarly, with the rotational Killing vector $\phi^a$, the angular momentum of the test particle is given by\cite{gao2013destroying}
\begin{equation}
\begin{aligned}
L&=(m u_{a}+q A_{a})\phi^{a}\\&=\int_{\Sigma}\left(\mu u_{a}+\rho A_{a}\right) \phi^{a}  \,^{3}\!\epsilon\\&
=-\int_{\Sigma}\left(u^{b}\Delta T_{ab}+u^{b}j_{b} A_{a}\right) \phi^{a}  \,^{3}\!\epsilon\\&=-\int_{\Sigma}\left(\Delta T_{ab}+j_{b} A_{a}\right)\phi^{a}u^{b} \,^{3}\!\epsilon.\label{dsfxc}
\end{aligned}
\end{equation}

After the test particle is thrown into the black hole, the Einstein's equation and Maxwell's equation are modified as
\bean
G_{ab}\eqn R_{ab}-\frac{1}{2}g_{ab}R=8\pi \left(T_{ab}^{EM}+\Delta T_{ab}\right),\label{gabgab} \\
\nabla^{a}F_{ab}\eqn-4\pi j_{b} \,.\label{maker}
\eean
From
\begin{equation}
\nabla^{a} T_{ab}^{EM}=-F_{bc}j^{c}\,,\label{msv}
\end{equation}
we obtain
\begin{equation}
 \nabla^{a}\left(\Delta T_{ab}\right)=F_{bc}j^{c}\label{dfce}.
\end{equation}
 Contracting  \eq{dfce} by any Killing field $\xi^{b}$ and using the Killing equation $\nabla_{\left(a\right.}\xi_{\left.b\right)}=0$, we can rewrite \eq{dfce} as
\begin{equation}
\nabla^{a}\left(\xi^{b} \Delta  T_{a b}\right)-\xi^{b} F_{b c} j^{c}=0\,.\label{asrg}
\end{equation}
By making use of $\mathscr{L}_{\xi}A=\xi\cdot dA+d(\xi\cdot A)=0$ \cite{sorce2017gedanken},  we have $\xi^{b} F_{b c}= \xi\cdot dA=-d\left(\xi\cdot A\right)$.
Then \eq{asrg} becomes
\begin{equation}
\nabla^{a}\left(\xi^{b}  \Delta  T_{a b}\right)+d(\xi\cdot A) j^{c}=\nabla^{a}\left(\xi^{b}  \Delta  T_{a b}\right)+\grad_c(\xi^b  A_b) j^{c}=0\label{adxc}.
\end{equation}
Together with the conservation of charge current
\begin{equation}
 \nabla_{c}j^{c}=0,\label{vcx}
\end{equation}
we obtain
\begin{equation}
\nabla^{b}\left(\xi^{a}\Delta  T_{a b}+\xi^{a} A_{a} j_{b}\right) =0.\label{asddg}
\end{equation}

\begin{figure}[!hbt]
  \centering
  \includegraphics[width=14cm,trim=-180 210 50 0,clip]{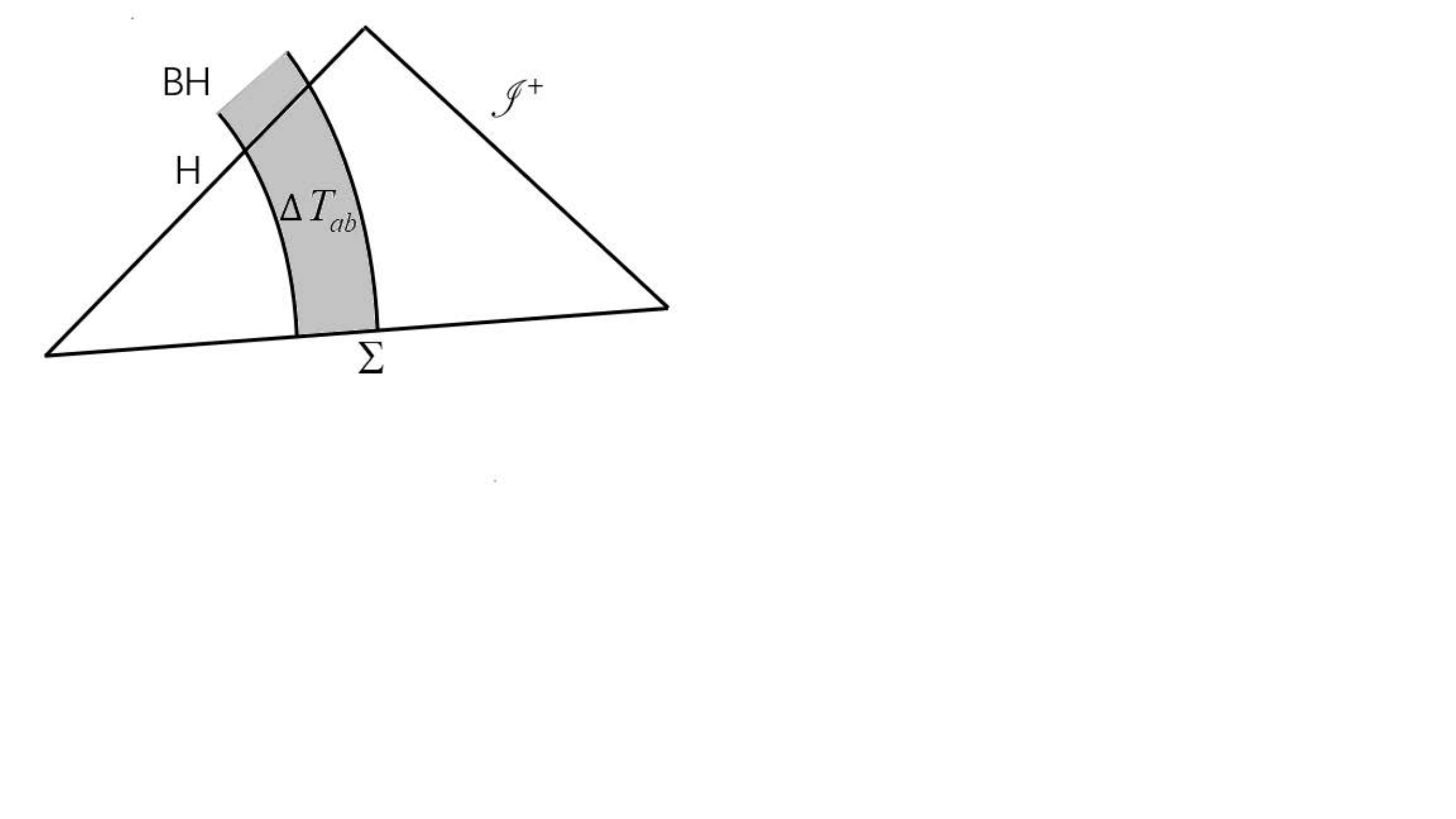}
  \caption{ A spacetime diagram representing charged matter falling into a black hole.}\label{fig5.1}
\end{figure}

Now, we choose a spacetime region bounded by the event horizon $H$, the spacelike hypersurface $\Sigma$, and the null infinity $\mathscr{I}^+$, as  shown in \fig{fig5.1}. Since we assume that all the matter falls into the black hole, we can apply Stokes' theorem to \eq{asddg} and obtain
\begin{equation}
\int_{\Sigma}\left(\xi_{a} \Delta  T^{a b}+\xi_{a} A^{a} j^{b}\right) \epsilon_{b d e f}+\int_{H}\left(\xi_{a} \Delta  T^{a b}+\xi_{a} A^{a} j^{b}\right) \epsilon_{b d e f}=0,\label{asdfa}
\end{equation}
where $\epsilon$ is the 4-volume element.

For any vector field $v^a$, we have \cite{wald1984general}
 \begin{equation}
  v^{a}\epsilon_{abcd}=n_a v^a\epsilon_{bcd}, \label{bskj}
 \end{equation}
 where $\epsilon_{bcd}$ is the  volume element induced by $n_a$, i.e. $\epsilon_{a b c d}=n_{a}\wedge\epsilon_{b c d}$. Therefore, \eq{asdfa} becomes
\begin{equation}
\int_{\Sigma}\left(\xi_{a} \Delta  T^{a b}+\xi_{a} A^{a} j^{b}\right) u_{b}\epsilon_{def}+\int_{H}\left(\xi_{a} \Delta  T^{a b}+\xi_{a} A^{a} j^{b}\right)    k_{b}\epsilon_{def}=0,\label{mscd}
\end{equation}
where  $k^{a}$ is the normal vector of  $H$.
By choosing  $\xi_{a}$ in \eq{mscd} as $t_a$ and $\phi_{a}$ respectively, \eqs{ases} and \meq{dsfxc} can be written as integrations on the horzion
\begin{equation}
E=-\int_{H}\left(t_{a} \Delta  T^{a b}+t_{a} A^{a} j^{b}\right) k_{b}\epsilon_{def},\label{uh}
\end{equation}
\begin{equation}
L=\int_{H}\left(\phi_{a} \Delta  T^{a b}+\phi_{a} A^{a} j^{b}\right) k_{b}\epsilon_{def}.\label{zx}
\end{equation}
Furthermore, \eqs{uh} and \meq{zx} can be expressed explicitly as
\begin{equation}
E=\int_{0}^{\infty}d V\int d^{2}S \left(\Delta T_{a b}+A_{a} j_{b}\right)t^{a}k^{b},\label{uhfd}
\end{equation}
\begin{equation}
L=-\int_{0}^{\infty}d V\int d^{2}S \left(\Delta T_{a b}+A_{a} j_{b}\right)\phi^{a}k^{b}\,,\label{zxsd}
\end{equation}
where $V$ is the affine parameter of the null geodesics with tangent $k^a$ and $S$ is the intersection of $\Sigma$ and $H$.

Similarly, applying Stokes' theorem to \eq{vcx}, we  obtain
\begin{equation}
\int_{\Sigma} j^{a}\epsilon_{abcd}+\int_{H} j^{a}\epsilon_{abcd}=0\,,\label{ogkj}
\end{equation}
which gives the expression of the charge
\begin{equation}
q=-\int_{0}^{\infty} d V \int d^{2} S j_{b}k^{b}.\label{qq}
\end{equation}

To derive the  ``physical process version" of the first law, we start with
 the Raychauduri equation
\begin{equation}
\frac{d\theta}{dV}=-\frac{1}{2}\theta^2-\sigma_{ab}\sigma^{ab}-R_{ab}k^{a}k^{b}\,.\label{asdfff}
\end{equation}
On the horizon, the quadratic terms $\theta^2$ and $\sigma_{ab}\sigma^{ab}$  are neglible. Then together with Einstein's equation, we have
\begin{equation}
\frac{d \theta}{d V}=-8 \pi \Delta T_{\mathrm{ab}} k^{a} k^{b}\,.\label{dcxsh}
\end{equation}

Let $\chi^{a}=t^{a}+\Omega_{H} \phi^{a}$ be the horizon Killing vector field, which is related to $k^a$ on the horizon by \cite{wald1994quantum}
\begin{equation}
\begin{aligned}
k^{a} &=\frac{1}{\kappa V} \chi^a.\label{opsd}
\end{aligned}
\end{equation}
In addition, on the right-hand side of \eq{dcxsh}, we can replace $k^a$ with \eq{opsd}.

Thus, multiplying both sides of \eq{dcxsh} by $\kappa V$ and integrating over the horizon, we obtain
\begin{equation}
\begin{aligned}
\kappa\int_{0}^{\infty} d V \int d^{2}S \quad  V \frac{d \theta}{d V} &=-8 \pi \int_{0}^{\infty} d V \int d^{2}S \quad \Delta T_{a b}\left(t^{a}+\Omega_{H} \phi^{a}\right) k^{b}.\label{bgf}
\end{aligned}
\end{equation}
In addition, subtracting and adding the term $8 \pi \int_{0}^{\infty} d V \int d^{2}S A_{a} j_{b} \chi^{a} k^{b}$ to the right-hand side of \eq{bgf}, we obtain
\begin{equation}
\begin{aligned}
\kappa\int_{0}^{\infty} d V \int d^{2}S \quad  V \frac{d \theta}{d V} &=-8 \pi \int_{0}^{\infty} d V \int d^{2} S\left(\Delta T_{a b}+A_{a}j_{b}\right)\left(t^{a}+\Omega_{H} \phi^{a}\right) k^{b}  \\
&+8 \pi \int_{0}^{\infty} d V \int d^{2}S\quad A_{a} j_{b} \chi^{a} k^{b}.\label{klnmd}
\end{aligned}
\end{equation}
The left side of \eq{klnmd} can be evaluated by integration by parts:
\bean
\kappa\int d^{2}S\int_{0}^{\infty} V \frac{d \theta}{d V} d V \eqn \kappa\int d^{2}S (\theta V)|^{\infty}_{0}-\kappa\int d^{2}S\int_{0}^{\infty} \theta d V.\label{cxdf}
\eean
The first term  on the right side of \eq{cxdf} vanishes since the black hole settles down to a stationary final state. According to $\theta=\frac{1}{A}\frac{dA}{dV}$, the second term on the right-hand side of \eq{cxdf} is just $-\kappa\Delta A$.

By making use of \eqs{uhfd}, \meq{zxsd}, \meq{qq}, and integrating \eq{klnmd} over the horizon, we obtain
\begin{equation}
-\kappa \Delta A =-8 \pi\left( E-\Omega_{H}L\right)+8 \pi q\Phi_{H},\label{lh}
\end{equation}
where $\Phi_{H}=-A_{a}\chi^{a}|_{H}$ is the electric potential on the horizon. \eq{lh} is the ``physical process version" of the first law for a charged axisymmetric  black hole.

Now, we apply this first law to the MKN black hole. Rewrite \eq{lh} as
\begin{equation}
 E=\frac{\kappa\Delta A}{8 \pi} +\Omega_{H}L+\Phi_{H}q\label{2dh}.
\end{equation}
Subtracting $\left(\Omega_{int}L+\Phi_{int}q\right)$ from both the left and right sides of \eq{2dh}, we obtain
\begin{equation}
E-\Omega_{int}L-\Phi_{int}q=\frac{\kappa\Delta A}{8 \pi} +\Omega_{H}L+\Phi_{H}q-\left(\Omega_{int}L+\Phi_{int}q\right)\label{hhjk},
\end{equation}
where $\Omega_{int}$ and $\Phi_{int}$ are constants introduced in \eq{innnn2}.
Multiplying both sides of the \eq{hhjk} are multiplied by a constant $\alpha$, we have
\begin{equation}
\alpha\left(E-\Omega_{int}L-\Phi_{int}q\right)=\alpha\left(\frac{\kappa\Delta A}{8 \pi} +\Omega_{H}L+\Phi_{H}q-\Omega_{int}L-\Phi_{int}q\right) \label{ccfhg}.
\end{equation}
Let
\begin{equation}
E'=\alpha\left( E-\Omega_{int}L-\Phi_{int}q\right),
\end{equation}
then \eq{ccfhg} becomes
\begin{equation}
E'=\alpha\left( \frac{\kappa\Delta A}{8 \pi}+(\Phi_{H}-\Phi_{int})q+(\Omega_{H}-\Omega_{int})L\right).\label{xmew}
\end{equation}
By comparing \eq{xmew} with \eq{innnn2}, we see that $E'$  is just $\delta \tilde M$, which answers how the black hole parameter changes after the particle enters the horizon.

\section{Validity of the second law of thermodynamics}\label{appc}
Recently, the second law of thermodynamics of some stationary black holes with the cosmological constant has been verified \cite{Zeng:2019aao,Zeng:2019huf,Zeng:2019hux}. In this section, we shall prove that the second law of thermodynamics is still valid  as the test particle falls into the magenetized black hole \cite{wald1994quantum}. As shown in appendix \ref{appb}, the left side of \eq{bgf} is $-\kappa\Delta A $.
Therefore, \eq{bgf} becomes
\bean
-\kappa\Delta A=-8 \pi \int_{0}^{\infty} d V \int d^{2}S \quad \Delta T_{a b}\left(t^{a}+\Omega_{H} \phi^{a}\right) k^{b}.\label{us}
\eean
 Substituting \eq{opsd} into the null energy condition $\Delta T_{\mathrm{ab}} k^{a} k^{b}\ge 0$,  we have
 \bean
  \Delta T_{a b}\left(t^{a}+\Omega_{H} \phi^{a}\right) k^{b}\ge 0,\label{hk}
 \eean
which gives immediately that
 $\Delta A\ge 0$.  By comparing \eq{xmew} with \eq{innnn2}, we see that $\Delta A$ is just $\delta A_{H}$ in \eq{innnn2}, i.e. $\delta A_{H}\ge 0$, which  indicates that the second law of thermodynamics still holds after we redefine the energy of the test particle.

\end{document}